\documentclass[fleqn,usenatbib]{mnras}
\usepackage{amsmath}
\usepackage{xcolor, graphicx}
\usepackage{amssymb}
\usepackage{soul}
\usepackage{txfonts}
\usepackage{caption}

\newcommand{\Msun}{\rm M_\odot}

\begin{document}

\title[Globular Cluster Dark Matter Connection]{The SLUGGS Survey: globular clusters and the dark matter content of early-type galaxies}
\author[D. A. Forbes et al.]{Duncan A. Forbes$^{1}$\thanks{E-mail:
dforbes@swin.edu.au}, Adebusola Alabi$^{1}$, Aaron J. Romanowsky$^{2,3}$, Jean P. Brodie$^{3}$, Jay Strader$^{4}$
\newauthor 
Christopher Usher$^{5}$, Vincenzo Pota$^{6}$
\\
$^{1}$Centre for Astrophysics \& Supercomputing, Swinburne
University, Hawthorn VIC 3122, Australia\\
$^{2}$Department of Physics and Astronomy, San Jos\'e State
University, One Washington Square, San Jose, CA 95192, USA\\
$^{3}$University of California Observatories, 1156 High Street, Santa Cruz, CA 95064, USA\\
$^{4}$Department of Physics and Astronomy, Michigan State
University, East Lansing, Michigan 48824, USA\\
$^{5}$Astrophysics Research Institute, Liverpool John Moores University, 146 Brownlow Hill, Liverpool L3 5RF, UK\\
$^{6}$INAF - Osservatorio Astronomico di Capodimonte, Salita Moiariello, 16, 80131 Napoli, Italy
}


\pagerange{\pageref{firstpage}--\pageref{lastpage}} \pubyear{2002}

\maketitle

\label{firstpage}

\begin{abstract}

A strong correlation exists between the total mass of a globular
cluster (GC) system and the virial halo mass of the host
galaxy. However, the total halo mass in this correlation is a
statistical measure conducted on spatial scales that are some ten
times that of a typical GC system.  Here we investigate the connection
between GC systems and galaxy's dark matter on comparable spatial
scales, using dynamical masses measured on a galaxy-by-galaxy
basis. Our sample consists of 
17 well-studied massive (stellar mass $\sim$10$^{11}$ M$_{\odot}$) 
early-type galaxies from
the SLUGGS survey.  We find the strongest correlation to be that of
the blue (metal-poor) GC subpopulation and the dark matter
content. This correlation implies that the dark matter mass of a
galaxy can be estimated to within a factor of two from careful imaging
of its GC system. The ratio of the GC system mass to that of the enclosed 
dark matter is nearly constant. 
We also find a strong correlation
between the fraction of blue GCs and the fraction of enclosed dark
matter, so that a typical galaxy with a blue GC fraction of 60 per
cent has a dark matter fraction of 86 per cent over similar spatial
scales.  Both halo growth and removal (via tidal stripping) may play
some role in shaping this trend. In the context of the two-phase model
for galaxy formation, we find galaxies with the highest fractions of
accreted stars to have higher dark matter fractions for a given
fraction of blue GCs.

\end{abstract}


\section{Introduction}

The globular cluster (GC) systems of galaxies, including the Milky Way and M31, reveal a remarkable
correlation with galaxy halo mass. 
The correlation has a slope consistent 
with unity so that the ratio of the total GC system
mass (M$_{GCS}$) to the total halo mass (M$_{h}$) is roughly constant
with a value of a few times 10$^{-5}$ over 5 orders of magnitude
(Blakeslee et al. 1997; 
Spitler et al. 2008; Spitler \& Forbes 2009; Georgiev et al. 2010;
Harris et al. 2013; Hudson et al. 2014; Durrell et al. 2014; Harris et
al. 2015). The scatter of the correlation is about 0.3 dex (e.g. 
Harris et al. 2015).

This near linear relation may simply reflect the gas
content of a given dark matter halo at the time of GC formation and is
close to the scaling predicted by Kravtsov \& Gnedin (2005) of
M$_{GCS} \propto M_{h}^{1.13\pm0.08}$ from their cosmological simulation.
Over cosmic time, it is expected that the ratio
M$_{GCS}$/M$_{h}$ is reduced   
due to GC
destruction (Katz \& Ricotti 2014). For further discussion of these issues 
we refer the reader to Harris et al. (2015).

The GC systems of large galaxies also reveal a bimodal colour distribution
that is well represented by blue and red subpopulations. This
bimodality is in turn thought to reflect differences in metallicity
(Brodie et al. 2012; 
Usher et al. 2012). Based on spatial distributions, metallicities and kinematics, 
previous GC studies have generally associated the
red/metal-rich subpopulation with the stellar bulge/elliptical component of
galaxies, and the blue/metal-poor subpopulation with the halo 
(Forte et al. 2007; Forbes et al. 2012; Pota et al. 2013; Forte et
al. 2014). In the currently popular two-phase scenario for early-type
galaxy formation (e.g. Oser et al. 2010), the halo is largely built up
by stars formed `ex-situ' in lower mass satellites which are later accreted.
However, as well as an accretion origin, GC system metallicity gradients
observed in a few well studied galaxies suggest that some blue GCs may
have formed `in-situ' along with the bulk of the red GCs 
(Harris 2009; Forbes et al. 2011).

It is thus important to investigate which GC subpopulation (i.e. blue
or red) correlates with the dark matter content of a galaxy. In the
largest study to date, Harris et al. (2015) split some 300 GC systems
into blue and red subpopulations, and examined the correlation with
halo mass (derived from a weak lensing study and a statistical
relationship between stellar and halo mass). They found the blue GCs
to have a slope close to unity, with the red GCs revealing a
significantly steeper slope. Both subpopulations revealed similar
scatter.

One limitation of past work is that GC systems, which have a
radial extent of about ten R$_e$ (the half-light radius of the 
galaxy starlight), have been
compared to halo masses which are measured on scales of the virial radius or $\sim$100
R$_e$. It is unclear whether the relation holds for 
spatial scales that are better matched 
(although see Kavelaars 1999 for an early 
investigation), and in particular if the red/blue GC fractions vary with
the enclosed dark matter fraction or the enclosed dark matter mass.

Such a connection between GCs and dark matter on 
comparable scales is plausible because
GCs are very old (e.g. Forbes et al. 2015) 
and formed during the early phases of collapse of
their host dark matter halos.  The GCs should therefore be preferentially
associated with the central regions of present-day dark matter halos
(Diemand et al. 2005), rather than with the total halo mass,
whose late-time growth may be dominated more by ``pseudo-evolution''
than by true infall (Diemer et al. 2013).

Here we use recent dynamical 
estimates of the dark matter content of individual 
early-type galaxies at radii comparable to their 
GC systems. In particular, we examine the dark matter on 
both a fixed spatial scale of 8 R$_e$ and on a scale that matches the
maximum extent of each GC system. 
Our sample consists of massive
early-type galaxies with high quality GC system imaging and estimates
of their enclosed dark matter content. Also for the first time, 
we use dynamical estimates of the dark matter mass rather than 
statistical scaling relations in the comparison with GC system mass. 

\section{Data}

The galaxy sample for this work comes from the SLUGGS survey 
(http://sluggs.swin.edu.au) of nearby
early-type galaxies (Brodie et al. 2014). The survey focuses on both
the galaxy starlight and the GC systems of the galaxies. The galaxies cover a 
range of environments from the field to central dominant cluster galaxies. 
All have stellar masses greater than 10.3 in the log, and so probe the 
high mass end of the U-shaped number of GCs per unit stellar mass 
relation (Forbes 2005; Peng et al. 2008). 

We take total (i.e. the pressure-supported mass plus the rotation mass) 
galaxy mass estimates 
from the recent work of Alabi
et al. (2016) who applied the tracer mass estimator (TME) eq. 26 of 
Watkins et al. (2010) 
to some 3500 GC radial velocities under the assumption of isotropic orbits. 
These velocities were obtained mostly with the
Keck telescope and the DEIMOS multi-object spectrograph.  
The TME gives galaxy mass estimates that are within a
factor of two compared to those published using X-ray gas, Planetary Nebulae (PNe), and 
stellar kinematics 
for the same galaxies at the equivalent radius. 
Here we include mass estimates from Alabi et al. 
for the 16 SLUGGS and 1 bonus galaxy  
for which the GC data all reach at least 8 R$_e$. 
The mass of each GC system is based on the number of GCs which is sourced mainly from the 
imaging compilation of Harris et al. (2013) supplemented by some more recent wide-field imaging studies
(Blom et al. 2012; Usher et al. 2013, Kartha et al. 2014, 2015). 
 The sample and data used in this work are summarised in Table 1.

\begin{table*}
\centering
{\small \caption{Galaxy sample and properties}}
\begin{tabular}{@{}lllllllllllll}
\hline
\hline
Galaxy & Dist. & R$_e$ & Env. & log~M$_{\ast}$ & GCS & $f_{\rm blue}$ & $N_{\rm GC}$ & f$_{DM}$ (8R$_e$)  & M$_{DM}$  (8R$_e$) & R$_{max}$ & f$_{DM}$ (R$_{max}$) & M$_{DM}$ (R$_{max}$) \\
$\rm [NGC]$ & [Mpc] 	& [kpc]  & & [$\Msun$] &  [kpc] & & & & [$10^{11} \Msun$] &  [R$_e$] &  & [$10^{11} \Msun$] 	\\
(1) & (2) & (3) & (4) & (5) & (6) & (7) & (8) & (9) & (10) & (11) & (12) & (13) \\
\hline
 720 & 26.9 & 4.56 & F & 11.35 & 78 &  0.64$\pm$0.05 &  1489$\pm$96      &                     0.65$\pm$0.09 &   3.88$\pm$1.22  &   19.1 & 0.81$\pm$0.05 &   9.60$\pm$2.02\\
 821 & 23.4 & 4.54 & F & 10.97 & 26 &  0.70$\pm$0.05 &   320$\pm$45	&		      0.85$\pm$0.04 &  4.90$\pm$0.97 &   8.7 & 0.85$\pm$0.04 &   4.92$\pm$1.00\\
1023 & 11.1 & 2.58 & G & 10.98 & 20 &  0.43$\pm$0.05 &   548$\pm$59	&		      0.63$\pm$0.08 &  1.52$\pm$0.32 &  16.2 & 0.76$\pm$0.05 &   2.99$\pm$0.52\\
1407 & 26.8 & 8.19 & G & 11.60 & 162 & 0.60$\pm$0.05 &  7000$\pm$2000	&		      0.82$\pm$0.04 & 16.36$\pm$1.66 &  14.1 & 0.89$\pm$0.02 &   32.76$\pm$2.69\\
2768 & 21.8 & 6.66 & G & 11.28 & 63 &  0.65$\pm$0.05 &   744$\pm$68	&		      0.83$\pm$0.05 &  8.75$\pm$1.38 &  11.4 & 0.83$\pm$0.05 &   9.05$\pm$1.45\\
3115 & 9.4 & 1.59 & F & 10.97 & 22$^{\dagger}$ & 0.52$\pm$0.04 & 550$\pm$80 &		      0.56$\pm$0.07 &  1.17$\pm$0.25 &  18.4 & 0.77$\pm$0.03 &   3.09$\pm$0.42\\
3377 & 10.9 & 1.90 & G & 10.44 & 27$^{\dagger}$ & 0.48$\pm$0.05 & 191$\pm$15	&	      0.69$\pm$0.06 &  0.59$\pm$0.12 &  14.3 & 0.78$\pm$0.04 &   0.97$\pm$0.16\\
3608 & 22.3 & 3.24 & G &10.82 & 43&  0.65$\pm$0.06 &   450$\pm$200 &		      0.83$\pm$0.09 &  3.08$\pm$1.03 &   9.8 & 0.85$\pm$0.09 &   3.56$\pm$1.18\\
4278 & 15.6 & 2.42 & G & 10.88 & 64 & 0.66$\pm$0.03  &  1378$\pm$194	&		      0.83$\pm$0.04 &  3.62$\pm$0.43 &  14.9 & 0.89$\pm$0.03 &   5.87$\pm$0.59\\
4365 & 23.1 & 5.94 & G & 11.48 & 133 & 0.43$\pm$0.01 &  6450$\pm$110	&		      0.84$\pm$0.03 & 14.37$\pm$1.65 &  12.9  & 0.90$\pm$0.02 &   25.27$\pm$2.56\\
4374 & 18.5 &  4.75 & C & 11.46 & 30 & 0.89$\pm$0.05 &  4301$\pm$1201	&		      0.89$\pm$0.04 &  20.72$\pm$5.31 &   9.2 & 0.89$\pm$0.03  &  22.82$\pm$5.50\\
4473 & 15.2 &  1.99 & C & 10.87 & 28$^{\dagger}$ &  0.57$\pm$0.05 &   376$\pm$ 97 &	      0.64$\pm$0.08 &  1.29$\pm$0.34 &  17.4 & 0.79$\pm$0.04  &  2.77$\pm$0.47\\
4486 & 16.7 &  6.56 & C & 11.53 & 92$^{\dagger}$ & 0.73$\pm$0.05 & 13000$\pm$800 &     0.90$\pm$0.01 &  29.53$\pm$2.16 &  30.5 & 0.97$\pm$0.01 &  112.83$\pm$6.43\\
4494 & 16.6 &  3.94 & G & 11.02 & 55$^{\dagger}$ & 0.72$\pm$0.04 & 392$\pm$49 &      0.34$\pm$0.13 &  0.50$\pm$0.21  &  8.5 & 0.36$\pm$0.12 &   0.56$\pm$0.21\\
4526 & 16.4 &  3.58 & C & 11.23 & 50$^{\dagger}$ & 0.62$\pm$0.10 & 388$\pm$117 &     0.54$\pm$0.15 &  1.89$\pm$0.41 &  12.1  & 0.58$\pm$0.10  &  2.35$\pm$0.56\\
4649 & 16.5 &  5.28 & C & 11.55 & 40 & 0.63$\pm$0.05 &  4000$\pm$500	&		      0.79$\pm$0.04 & 12.75$\pm$1.11 &  24.3 & 0.93$\pm$0.01 &   46.24$\pm$3.53\\
\hline
3607 & 22.2 & 4.20 & G & 11.29 & 61 & 0.45$\pm$0.06 &   600$\pm$200	&	      0.66$\pm$0.18 &  3.64$\pm$1.41 &  20.7 & 0.81$\pm$0.08 &   8.35$\pm$2.45\\
\hline
\end{tabular}
\begin{flushleft}
{\small 
Notes: columns are (1) galaxy name, (2) distance, (3) effective radius and (4) 
environment (Field, Group or Cluster) from Brodie et al. (2014), (5) 
log total stellar mass from Forbes et al. (2015) 
with an assumed error of $\pm$0.2 dex (to allow for reasonable 
variations in IMF, age and metallicity),  
(6) globular cluster system extent from Kartha et al. (2015) or ${\dagger}$ 
assumed to be 14$\times$ the galaxy effective radius, 
(7) fraction of blue 
globular clusters and (8) total number of globular clusters from the compilation of  
Harris et al. (2013), except where we have used more recent wide-field imaging studies, i.e. 
NGC 720, 1023 and 2768  (Kartha et al. 2014), NGC 4278 (Usher et al. 2013), NGC 4365 (Blom et al. 2012), and for the blue GC 
fractions for NGC 3607 and NGC 3608 we use Kartha et al. (2015).  We also note that there was a typo in Foster et al. (2011) so the 
blue and red GC numbers were swapped by mistake. Additional columns are 
(9) dark matter fraction and (10) dark matter mass within 8 R$_e$, (11) maximum radius of the GC system in galaxy 
effective radii,  
(12) dark matter fraction and (13) dark matter mass within R$_{max}$  from Alabi et al. (2015). 
NGC 3607 is a SLUGGS bonus galaxy. }
\end{flushleft}
\end{table*}

\section{Results and Discussion}

The extent of a GC system scales with the mass and size of the host
galaxy. In particular, they typically extend to about 14$\times$ the
galaxy effective radius (Kartha et al. 2015). Previously, GC system
numbers, or inferred total GC system masses, have been compared to
virial masses which are measured on scales of $\sim$100 R$_e$. Here, we first
probe the galaxy dark matter within a fixed 8 R$_e$, then second, we
probe the galaxy dark matter at the maximum radius for which GC
velocities and mass estimates are available. The latter naturally
probes each individual GC system on a spatial scale that is closely
matched to the imaging (from which GC mass estimates are derived).

\begin{figure}
        \includegraphics[width=0.48\textwidth]{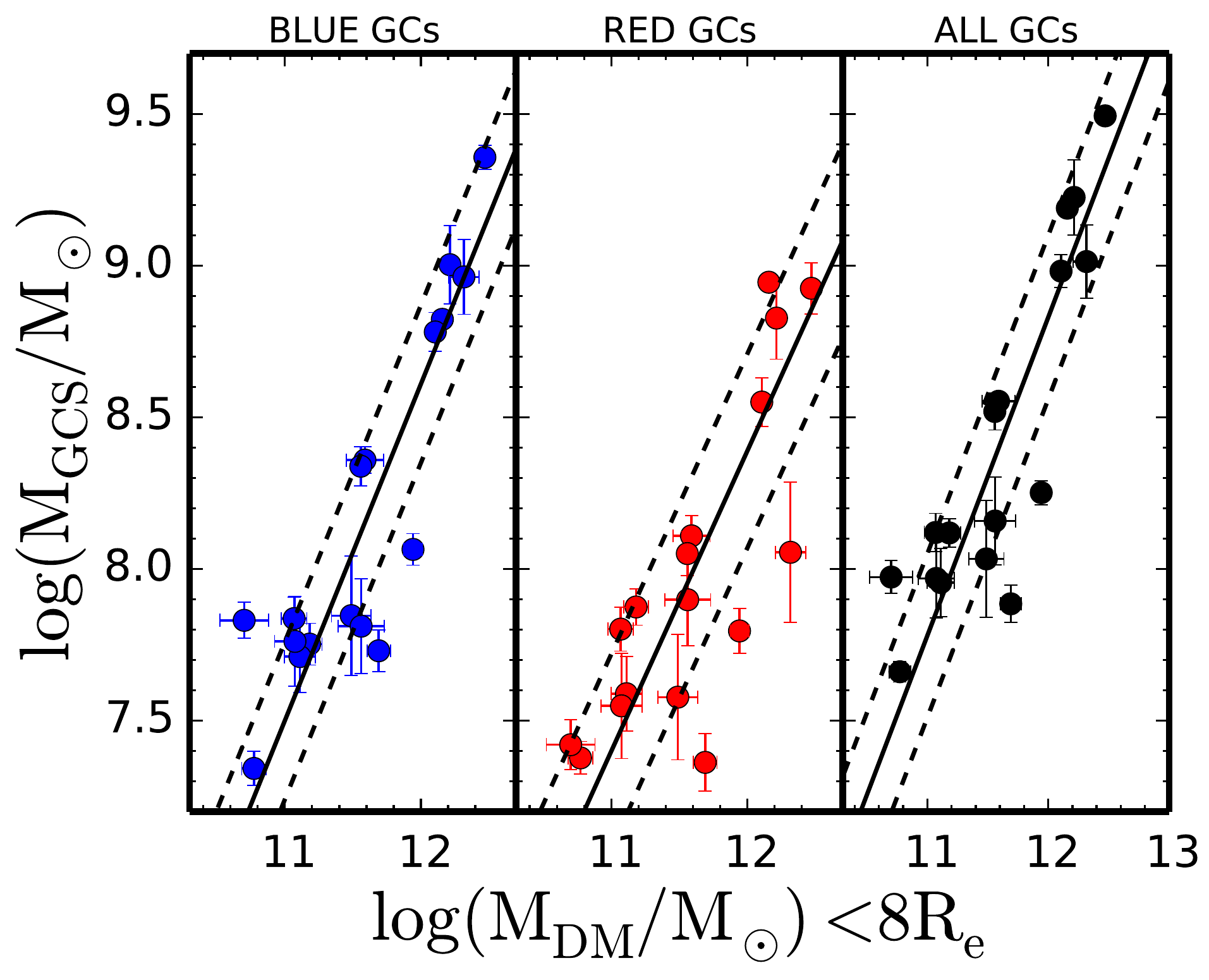}
	    \caption{\label{fig:corrGC2} Globular cluster system mass versus dark
 matter mass within 8 R$_e$.
The total mass contained in the
	    blue, red and all GCs is calculated 
assuming a constant mean GC mass. Solid and dashed lines show the best fit and 1$\sigma$ scatter. 
The blue, red and all GCs 
	    correlate with the enclosed dark matter content,  with the blue subpopulation
	    having the tightest relation. } 

\end{figure}


To calculate the total mass
in each GC subpopulation, we start with an estimate of the total number
of GCs and assign the appropriate fraction to each subpopulation based
on the literature studies (see Table 1). We then use a mean GC mass of
$2.4 \times 10^5 ~\Msun$ as used by Spitler et al. (2008) and Durrell
et al. (2014) to estimate the mass in each GC
subpopulation. Variations of the mean mass with subpopulation will
only change the normalisation of Fig. 1. Similarly, variations of the
mean mass with galaxy magnitude are not included due to the small
magnitude range of our sample. 

The dark matter mass (M$_{DM}$) 
within 8 R$_e$ (as listed in Table 1) is 
calculated from the total mass (M$_{\rm tot}$) minus 
the stellar mass (M$_{\ast}$) within 8 R$_e$ (see Alabi et al. 2016 for details). 
We note that the dark matter mass contained within 8 R$_e$ 
is $\sim$10 per cent of the total virial 
dark matter mass for a massive halo with an NFW-like profile. 

Fig. 1 shows that the mass of the blue, red and all GCs have a
strong correlation 
with the dark matter content within 8 R$_e$. 
A similar finding was found by Harris et al. (2015) who used 
the total halo virial mass.
Our results from a linear fit to the data are given in Table 2. Here we 
use a generalised chi-squared minimisation fit as given in the R package 
called hyper.fit. A weighted fit is performed on both coordinates based on their uncertainities, 
with an assumption 
that their uncertainties are independent. All data points were included in the fit.


Like Harris et al., we find less scatter for the blue GCs compared 
to the red ones, but not a statistically significant difference. We have also examined the 
correlation of GC masses with total (i.e. dark plus stellar) mass and find the relations to have 
slightly more scatter than using the dark matter mass within R$_{max}$. 

%

\begin{table}
\centering
{\small \caption{Fits to globular cluster system mass with dark matter}}
\begin{tabular}{@{}rlll}
\hline
 & Slope & Intercept & rms\\
\hline
BGC--DM within 8 R$_e$  & 1.11$\pm$0.14 & --4.71$\pm$1.59 & 0.26$\pm$0.06\\
RGC--DM within 8 R$_e$  & 0.99$\pm$0.17& --3.49$\pm$1.98 & 0.32$\pm$0.07\\
GC--DM within 8 R$_e$  & 1.05$\pm$0.14 & --3.77$\pm$1.63 & 0.27$\pm$0.05\\
\hline
BGC--DM within R$_{max}$ & 0.98$\pm$0.09 & --3.43$\pm$1.11 & 0.20$\pm$0.05\\
RGC--DM within R$_{max}$ & 0.88$\pm$0.11 & --2.47$\pm$1.35 & 0.25$\pm$0.05\\
GC--DM within R$_{max}$ & 0.93$\pm$0.09 & --2.61$\pm$1.09 & 0.20$\pm$0.04\\
\hline
\end{tabular}
\begin{flushleft}
{\small 
Notes: BGC = blue globular clusters, RGC = red globular clusters, GC = total globular cluster system. 
}
\end{flushleft}
\end{table}


 
 

The best fit to the mass of the overall GC system versus the enclosed 
dark matter content within 8 R$_e$  implies a 
near constant ratio of M$_{GCS}$/M$_{DM}$ $\approx$ 6.5 $\times$ 10$^{-4}$. 
An estimate of the expected value can be found  by first estimating the total mass 
of stars formed in a globular cluster system M$_{GCS}$ for a given dark matter mass M$_{DM}$ (Hudson et al. 
2014): 

\begin{equation}
\frac{M_{GCS}}{M_{DM}} = \frac{M_{GCS}} {M_{\ast}} \times \frac{M_{\ast}} {M_{b}} \times \frac{\Omega{_b}}{\Omega{_m}}
\end{equation}

Here $\Omega_b/ \Omega_m$ is the universal baryon fraction of
0.17. The ratio M$_{GCS}$/M$_{\ast}$ is the fraction of mass in the
overall GC system relative to the galaxy stellar mass. For a typical
log M$_{\ast}$ = 11.2 galaxy, Table 2 gives the ratio to be $\sim$2$\times
10^{-3}$ (i.e. similar to the value found by McLaughlin (1999) for massive
galaxies). M$^{\ast} / M_{b}$ is the efficiency of forming stars for a
given available baryonic mass. Based on SDSS observations, Guo et
al. (2010) showed in their figure 5 that this ratio is about 0.1 for
massive galaxies of log M$_{h}$ $\approx$ 13. Thus equation 7 becomes:

\begin{equation}
\frac{M_{GCS}}{M_{h}} = 2 \times 10^{-3} \times 0.1 \times 0.17 = 3.4 \times 10^{-5}
\end{equation}

This number is close to the values found in the literature when
comparing the overall GC system mass with the virial halo mass as
summarised by Harris et al. (2015). For example, Durrell et al. (2014), 
who used the same mean GC mass as we do, measured 2.9 $\times
10^{-5}$. In order to compare with the value we obtain from the fit to
Fig. 1 we need to divide by the fraction of the total halo mass
contained within 8 R$_e$. Assuming an NFW-like profile for a massive
halo, the fraction of the total halo mass within 8 R$_e$ is 10 per cent. This implies an expected ratio of
M$_{GCS}$/M$_{DM}$ within 8 R$_e$ of 3.4 $\times 10^{-4}$. Given the simplified assumptions, this is comparable
to our measured value of 6.5 $\times 10^{-4}$.

In Fig. 1 we explored the correlation of GC system mass with the dark
matter content within a fixed radius, i.e. 8 R$_e$. Now in Fig. 2 we
compare the blue, red and overall GC system mass to the dark matter
mass within the maximum radius for each GC system (R$_{max}$). In this case 
we subtract 100 per cent of the stellar mass from the total mass. 
Although R$_{max}$ varies from galaxy to galaxy (i.e. from 8.5 to 30
R$_e$), it is a better match to the spatial scale of each individual
GC system. The linear fit parameters are given in Table 2.




\begin{figure}
        \includegraphics[width=0.48\textwidth]{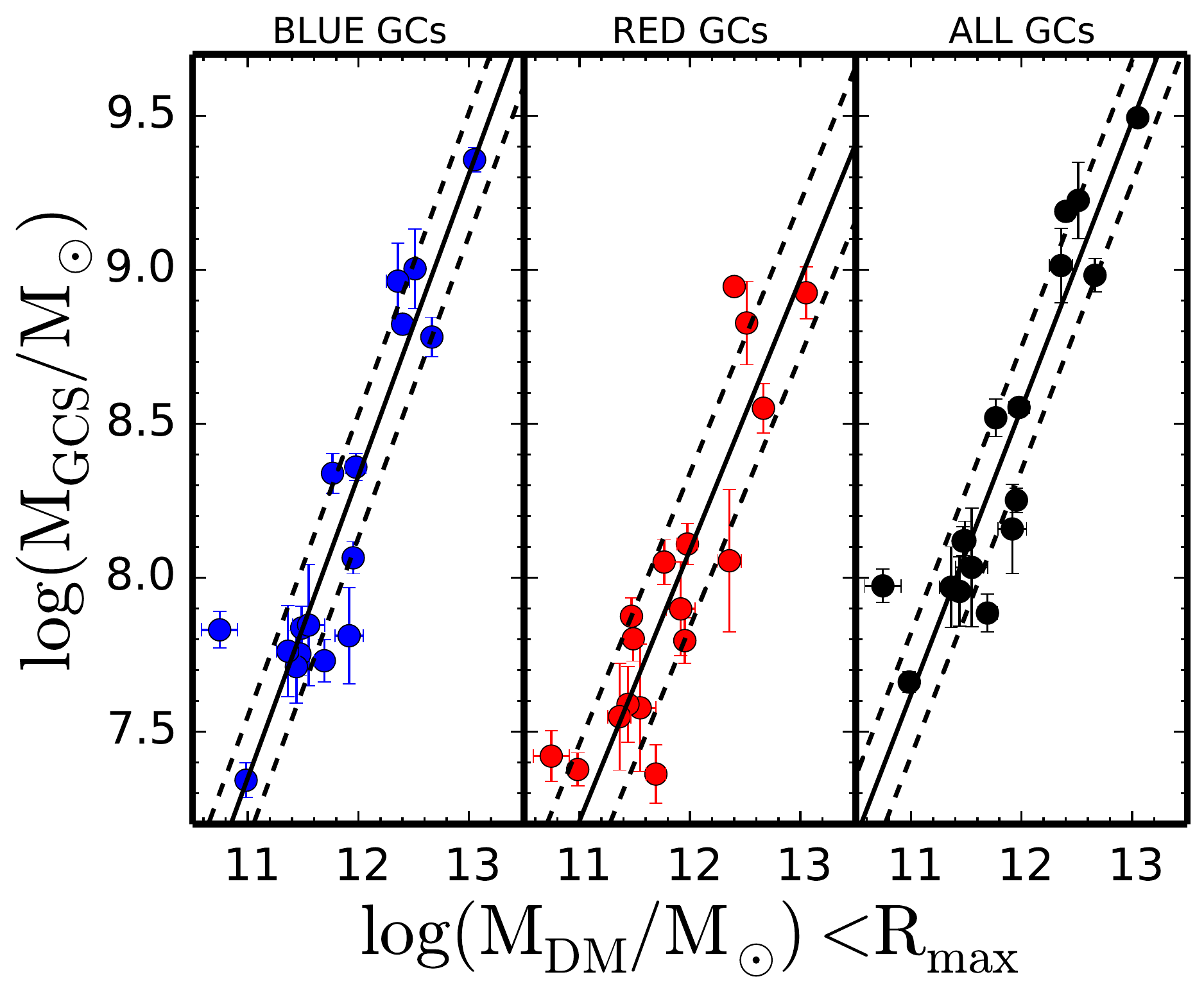}
	    \caption{\label{fig:corrGC2} Globular cluster system mass versus dark matter
mass within R$_{max}$.
The total mass contained in the
	    blue, red GCs and all GCs is calculated assuming a constant mean GC mass.
Solid and dashed lines show the best fit and 1$\sigma$ scatter. 	    
The blue, red and all GCs 
	    correlate with the stellar mass,  with the blue subpopulation
	    having the tightest relation. } 

\end{figure} 

Again, the blue subpopulation reveals the tightest correlation but we also find that the scatter is significantly reduced 
from 0.26 to 0.20 dex when changing from a fixed spatial scale of 8 R$_e$ to a matched spatial scale of R$_{max}$. 
This suggests that the mass in a GC system is strongly related to the enclosed dark matter mass. 
We have also examined the 
correlation of GC masses with total (i.e. dark plus stellar) mass and find the relations to have 
slightly more scatter than using the dark matter mass within R$_{max}$.

In Fig. 3 we show, for the first time, the metal--poor (blue) GC
subpopulation fraction ($f_{\rm blue}$) versus the fraction of dark
matter ($f_{\rm DM}$) enclosed within R$_{max}$, where $f_{\rm DM} =
M_{\rm DM}/M_{\rm tot}$.  
There is a strong correlation between $f_{\rm blue}$ and
$f_{\rm DM}$. The best fit to the data, excluding NGC 4494 and NGC
4526, is given in Table 3. Thus galaxies with GC systems that are
dominated by blue GCs also have mass distributions that are dominated
by dark matter.

None of the galaxies in the plot have blue GC fractions less than
0.4. The much larger sample of Harris et al. (2015) also reveals this
to be the case (with one exception, for which the imaging does not
cover the full GC system; W. Harris 2015, priv. comm.)  Thus
f$_{blue}$ $>$ 0.4 may represent a minimum, or floor, in the blue GC
fraction of galaxies. In contrast, the red GC fraction is known to be
zero in some lower mass galaxies. From the best fit relation,
f$_{blue}$ $>$ 0.4 corresponds to a lower limit in the enclosed dark
matter fraction of f$_{DM}$ of 0.79. A typical blue GC fraction of 0.6
corresponds to an enclosed dark matter fraction of 0.86.

\begin{figure}
\includegraphics[width=0.48\textwidth]{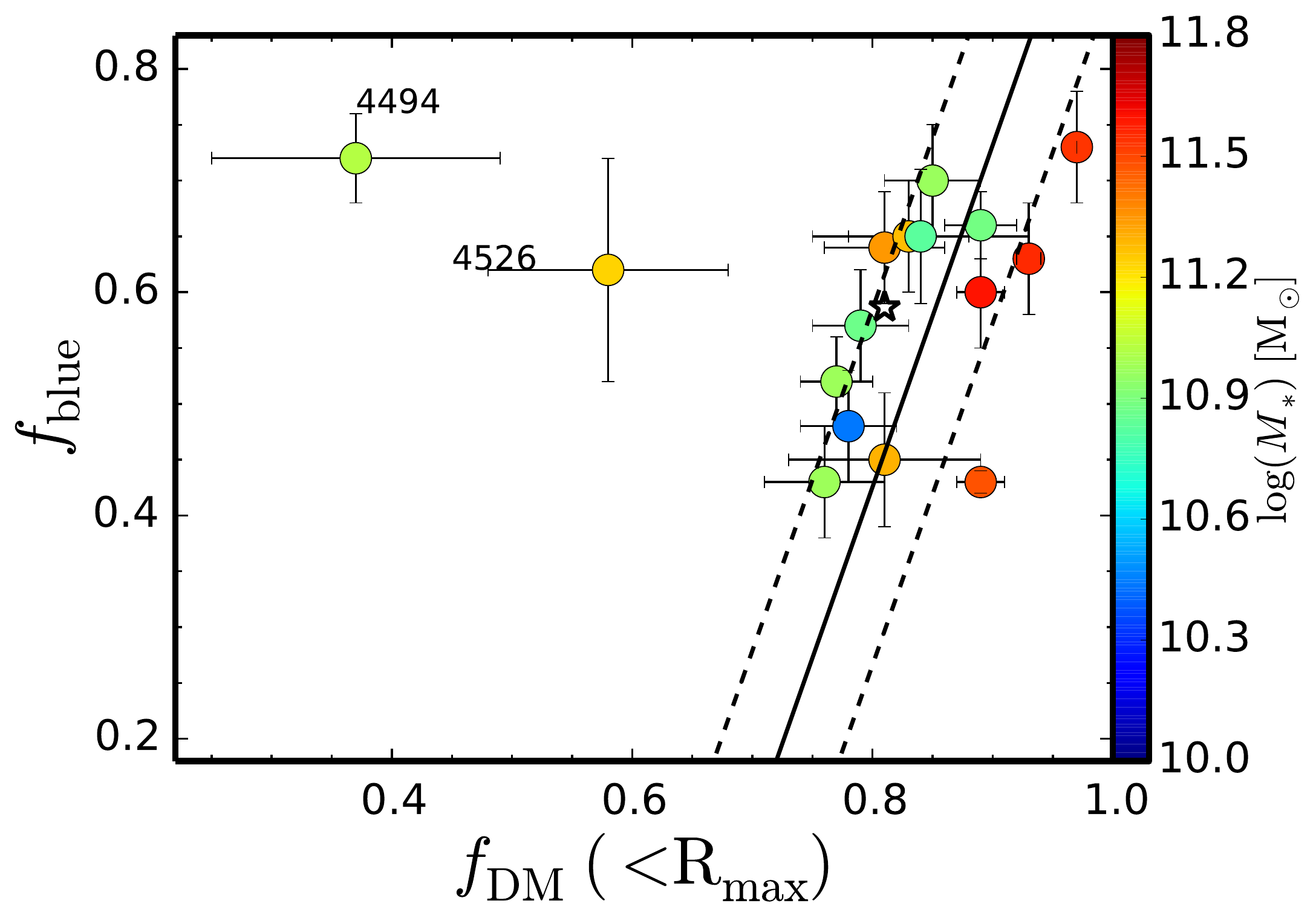}
\caption{Blue GC fraction versus enclosed dark matter fraction within R$_{max}$. 
The solid line is the best fit 
to the data (excluding NGC 4494 and NGC 4526), with dashed lines indicating 
	    1~$\sigma$ scatter. 
Points are colour-coded by stellar mass.
	    The open star indicates the location for a typical high mass galaxy from the 
	    study of Harris et al. (2015). 
A tight correlation between the blue GC and dark matter 
fraction is seen. 
The highest stellar mass galaxies tend to have the highest dark matter fraction for a given blue GC fraction. 
}
\end{figure} 


\begin{table}
\centering
{\small \caption{Fit to blue globular cluster fraction with dark matter fraction (excluding NGC 4494 and NGC 4526)}}
\begin{tabular}{@{}llll}
\hline
 & Slope & Intercept & rms\\
\hline
f$_{blue}$--f$_{DM}$ within R$_{max}$ & 3.06$\pm$1.28 & --2.02$\pm$1.11 & 0.16$\pm$0.07\\
\hline
\end{tabular}
\begin{flushleft}
\end{flushleft}
\end{table}

We also show in Fig.  3 the predicted blue GC fraction based on the
findings of Harris et al. (2015). First, we estimate the halo mass for
each galaxy in our sample using the M$_{\ast}$--M$_{h}$ relation
of Behroozi et al. (2010). Next, we use the f$_{blue}$--M$_{h }$ 
relation from Harris et al. to predict the blue GC fractions for our
galaxies,
 given their halo mass (using eq. 3 of Harris et al.).  In Fig.  3 we
 show this average predicted blue GC fraction versus the average
 enclosed dark matter fraction we measure for our sample (given by a star symbol). 
The predicted blue GC fraction agrees nicely with our empirical trend.
The trend in Fig.3 also appears to be consistent with the GC systems of dwarf galaxies. Such galaxies 
are  thought to be strongly dark matter dominated at large radii, and have blue GC fractions of 
approaching unity, i.e. they often contain no red GCs (Peng et al. 2008).

We now discuss the two deviant galaxies in Fig. 3 and whether their location in the plots
might be spurious. 
Two effects may explain the deviation of NGC 4526 from
the general trend. Firstly, it has the highest correction in our
sample for kinematic substructure (Alabi et al. 2016).
Uncorrected it would have $f_{\rm
DM}$ ${\sim}0.7$.  It also had a fairly large correction applied by
Peng et al. (2008) as the GC system is not fully covered by the ACS
field-of-view, i.e. uncorrected $f_{\rm blue}$ = 0.48. If
either of these corrections were not applied
then NGC 4526 would be consistent within the errors with the general
trend. Thus we suspect that NGC 4526 is not a true outlier and will obey the general trend with 
improved data.

NGC 4494 deviates most strongly from the general trend. Given that there appears to 
be a minimum, or floor, in the blue GC fraction of galaxies of f$_{blue}$ $>$ 0.4, 
changes to the blue GC fraction will not place NGC 4494 on the general trend. So we turn our 
attention to its low dark matter fraction. 
Indeed we are not the first to find a low fraction of dark matter. 
Several studies using PNe to model the mass of NGC 4494 
have derived a remarkably low dark matter content of $<$0.5 within 5 R$_e$ (Romanowsky et
al. 2003; Napolitano et al. 2009; Deason et al. 2012)  However, the more recent NMAGIC modelling 
of the extended stellar kinematics (Foster et al. 2011) and the PNe data by Morganti et al. 
(2013) found a higher dark matter fraction of f$_{DM}$ = 0.6 $\pm$ 0.1 within 5 R$_e$.
They assumed a logarithmic potential for the dark matter 
rather than an NFW-type profile, which leads to a higher dark matter mass and a relatively low stellar mass-to-light ratio. 
The wide range of dark matter fractions in the literature suggest that NGC 4494 is difficult to model dynamically, and 
future mass modelling should attempt to incorporate stellar, PNe and GCs velocity information simultaneously.  
If NGC 4494 follows the general trends for blue GC total mass (Figures 1 and 2) and blue GC fraction (Figure 3) found in this work, then 
a significant dark mass and dark matter fraction is expected.


A possible explanation for the overall correlation seen in Fig.  3 is the process of
halo growth via accretion of satellites. As
galaxies accrete low-mass satellites they will also accrete dark
matter and (preferentially) blue GCs. This tends to move a galaxy up
and to the right in the plot.  
The data points in Fig. 3 have been coded by their stellar mass. In the two-phase model for galaxy formation 
(e.g. Oser et al. 2010), the fraction of accreted (ex-situ) stars is a strong function of stellar mass. The plot shows 
a weak secondary trend for the highest stellar mass galaxies to have the highest dark matter fraction for a given blue GC fraction. 
This suggests that the accretion of many low mass satellite galaxies gives rise to a high ratio of dark 
matter to blue GCs in the outer regions of the most massive galaxies (Amorisco 2015).

Alternatively, galaxies may move down and to the left in
this plot via tidal stripping. For a typical galaxy in our sample 
the 2D surface density profile of a GC system
has a power-law slope of --2. Assuming spherical symmetry this
corresponds to --3 in 3D. At large radii an NFW profile slope is also
around --3. Thus both the GC radial profile and that of the dark
matter may follow a similar slope. This suggests that once tidal
stripping has removed the outermost dark matter, further stripping
would remove blue GCs (as they are more radially extended than the red
subpopulation) and dark matter in roughly equal proportion. Tidal
stripping of a galaxy with an initially high dark matter and blue GC
fraction would move it down and parallel to the trend seen in Fig. 3.  
Thus both tidal stripping and halo growth via accretion may play a role in the 
correlation of blue GC fraction and enclosed dark matter fraction.

\section{Conclusions}

Using a sample of well-studied massive early-type galaxies, we compare
the mass of their GC systems to the host galaxy's dark matter content on
similar spatial scales for the first time. We find a strong 
correlation between the mass of the blue (metal-poor) GC subpopulation and the
dark matter content within a fixed 8 R$_e$. An even stronger correlation is found when 
the radius for calculating the dark matter mass is matched to the spatial scale of the GC system.
Thus from careful imaging of a GC
system, one can infer the enclosed dark matter mass to within a factor of
two. 
The ratio of the 
GC system mass  to that of the enclosed dark matter is nearly constant, 
and this can be understood in terms of the universal
baryon fraction, the efficiency of GC formation and the radial profile
of dark matter. We also find a tight correlation 
between the fraction of blue GCs and the enclosed dark matter fraction. 
GC systems appear to have a minimum blue fraction of 40 per cent, although 
a typical galaxy, with blue GC fraction of 60 per cent, 
has 86 per cent of its mass in
dark matter (and 14 per cent in stars). 
The dark matter mass of NGC 4494 has been debated in the literature, but based on our findings we support 
claims that it is heavy with dark matter.
Both halo growth and removal (via tidal stripping) may play some 
role in shaping this trend. We suggest that galaxies with the highest accretion fractions reveal 
higher dark matter fractions for a given blue GC fraction. 

\section{Acknowledgements}

We thank J. Janz for useful discussion and the referee, Bill Harris, for 
his helpful suggestions. 
DAF thanks the ARC for financial support via DP130100388. This work was 
supported by NSF grant AST-1211995.

\section{References}

Alabi A., et al. 2016, in prep.\\
Amorisco, N., 2015, arXiv:1511.08806\\
Behroozi P.~S., Conroy C., Wechsler R.~H., 2010, ApJ, 717, 379 \\
Blakeslee J.~P., Tonry J.~L., Metzger M.~R., 1997, AJ, 114, 482\\
Blom C., Spitler L.~R., Forbes D.~A., 2012, MNRAS, 420, 37\\
Brodie J.~P., Usher C., Conroy C., Strader J., Arnold J.~A., Forbes D.~A., 
Romanowsky A.~J., 2012, ApJ, 759, L33 \\
Brodie J.~P., et al., 2014, ApJ, 796, 52 \\
Diemand J., Madau P., Moore B., 2005, MNRAS, 364, 367 \\
Diemer B., More S., Kravtsov A.~V., 2013, ApJ, 766, 25\\
Durrell P.~R., et al., 2014, ApJ, 794, 103\\
Faifer F.~R., et al., 2011, MNRAS, 416, 155 \\
Forbes D.~A., 2005, ApJ, 635, L137\\
Forbes D.~A., Spitler L.~R., Strader J., Romanowsky A.~J., Brodie J.~P., 
Foster C., 2011, MNRAS, 413, 2943 \\
Forbes D.~A., Ponman T., O'Sullivan E., 2012, MNRAS, 425, 66 \\
Forbes D.~A., Pastorello N., Romanowsky A.~J., Usher C., Brodie J.~P., 
Strader J., 2015, MNRAS, 452, 1045\\
Forte J.~C., Faifer F., Geisler D., 2007, MNRAS, 382, 1947 \\
Forte J.~C., Vega E.~I., Faifer F.~R., Smith Castelli A.~V., Escudero C., 
Gonz{\'a}lez N.~M., Sesto L., 2014, MNRAS, 441, 1391\\
Foster C., et al., 2011, MNRAS, 415, 3393 \\
Georgiev I.~Y., Puzia T.~H., Goudfrooij  P., Hilker M., 2010, MNRAS, 406, 1967\\
Guo Q., et al., 2010, MNRAS, 404, 1111 \\
Harris W.~E., 2009, ApJ, 699, 254 \\
Harris W.~E., Harris G.~L.~H., Alessi M., 2013, ApJ, 772, 82\\
Harris W.~E., Harris  G.~L., Hudson M.~J., 2015, ApJ, 806, 36\\
Hudson M.~J., Harris G.~L., Harris W.~E., 2014, ApJ, 787, L5 \\
Kartha S.~S., Forbes D.~A., Spitler L.~R., Romanowsky A.~J., Arnold J.~A., 
Brodie J.~P., 2014, MNRAS, 437, 273 \\
Kartha S., et al. 2015, MNRAS, submitted\\
Katz H., Ricotti M., 2014, MNRAS, 444, 2377 \\
Kavelaars J.~J., 1999, ASPC, 182, 437\\ 
Kravtsov A.~V., Gnedin O.~Y., 2005, ApJ, 623, 650\\
McLaughlin D.~E., 1999, AJ, 117, 2398 \\
Morganti L., Gerhard O., Coccato L., 
Martinez-Valpuesta I., Arnaboldi M., 2013, MNRAS, 431, 3570 \\
Napolitano N.~R., et al., 2009, MNRAS, 393, 329 \\
Oser L., Ostriker J.~P., Naab T., Johansson P.~H., Burkert A., 2010, ApJ, 
725, 2312 \\
Peng  E.~W., et al., 2008, ApJ, 681, 197\\
Pota V., et al., 2013, MNRAS, 428, 389 \\
Romanowsky A.~J., Douglas N.~G., Arnaboldi 
M., Kuijken K., Merrifield M.~R., Napolitano N.~R., Capaccioli M., Freeman 
K.~C., 2003, Sci, 301, 1696 \\
Spitler L.~R., Forbes D.~A., Strader J., 
Brodie J.~P., Gallagher J.~S., 2008, MNRAS, 385, 361 \\
Spitler L.~R., Forbes D.~A., 2009, MNRAS, 392, L1\\
Usher C., et al., 2012, MNRAS, 426, 1475 \\
Usher C., Forbes D.~A., Spitler L.~R., Brodie J.~P., Romanowsky A.~J., 
Strader J., Woodley K.~A., 2013, MNRAS, 436, 1172 \\
Watkins L.~L., Evans N.~W., An J.~H., 2010, MNRAS, 406, 264 \\

\end{document}